\documentclass{article} 
\usepackage{iclr2025_conference,times}

\newcommand\changed[1]{\textcolor{black}{#1}}

\usepackage{amsmath,amsfonts,bm}

\def\eqref#1{equation~\ref{#1}}

\def\1{\bm{1}}

\DeclareMathAlphabet{\mathsfit}{\encodingdefault}{\sfdefault}{m}{sl}
\SetMathAlphabet{\mathsfit}{bold}{\encodingdefault}{\sfdefault}{bx}{n}

\usepackage{hyperref}
\usepackage{url}
\usepackage{booktabs}
\usepackage{multirow}
\usepackage{graphicx}
\usepackage{makecell}
\usepackage{wrapfig}
\usepackage{caption}

\title{Exploring Persona-dependent LLM Alignment for the Moral Machine Experiment}

\author{Jiseon Kim$^{1}$\thanks{Equal contribution.} , Jea Kwon$^{2*}$, Luiz Felipe Vecchietti$^{2*}$, Alice Oh$^{1}$, Meeyoung Cha$^{1,2}$
\\ $^1$Korea Advanced Institute of Science \& Technology (KAIST), Daejeon, South Korea  \\$^2$Max Planck Institute for Security and Privacy (MPI-SP), Bochum, Germany \\ \vspace{-4mm}
\\ \texttt{jiseon\_kim@kaist.ac.kr}\\ 
\vspace{-4mm}
\\ \texttt{\{jea.kwon,felipe.vecchietti,mia.cha\}@mpi-sp.org}\\
\texttt{alice.oh@kaist.edu}
}

\iclrfinalcopy 
\begin{document}

\maketitle

\begin{abstract}

Deploying large language models (LLMs) \changed{with agency} in real-world applications raises critical questions about how these models will behave. In particular, how will their decisions align with humans when faced with moral dilemmas? This study examines the alignment between LLM-driven decisions and human judgment in various contexts of the moral machine experiment, including personas reflecting different sociodemographics. We find that the moral decisions of LLMs vary substantially by persona, showing 
\changed{greater shifts in moral decisions for critical tasks than humans.} 
Our data also indicate an interesting partisan sorting phenomenon, where political persona predominates the direction and degree of LLM decisions.
We discuss the ethical implications and risks associated with deploying these models in applications that involve moral decisions.

\end{abstract}

\section{Introduction}
\label{intro}




Large language models (LLMs) demonstrate remarkable capabilities in various natural language processing tasks \changed{such as dialogue systems \citep{achiam2023gpt}, reasoning \citep{guo2025deepseek}, and robotic applications \citep{zeng2023large}}. 
Despite these advances, deploying LLMs in real-world applications for mission-critical tasks and sensitive scenarios, such as a key component of autonomous vehicles and ethical decision-support systems, raises urgent questions about their potential risks and robustness.
In particular, it is important to understand how these models make ethical decisions in moral dilemma scenarios and whether they align with human values.

\begin{figure}[!htpb]
\includegraphics[width=0.6\textwidth]{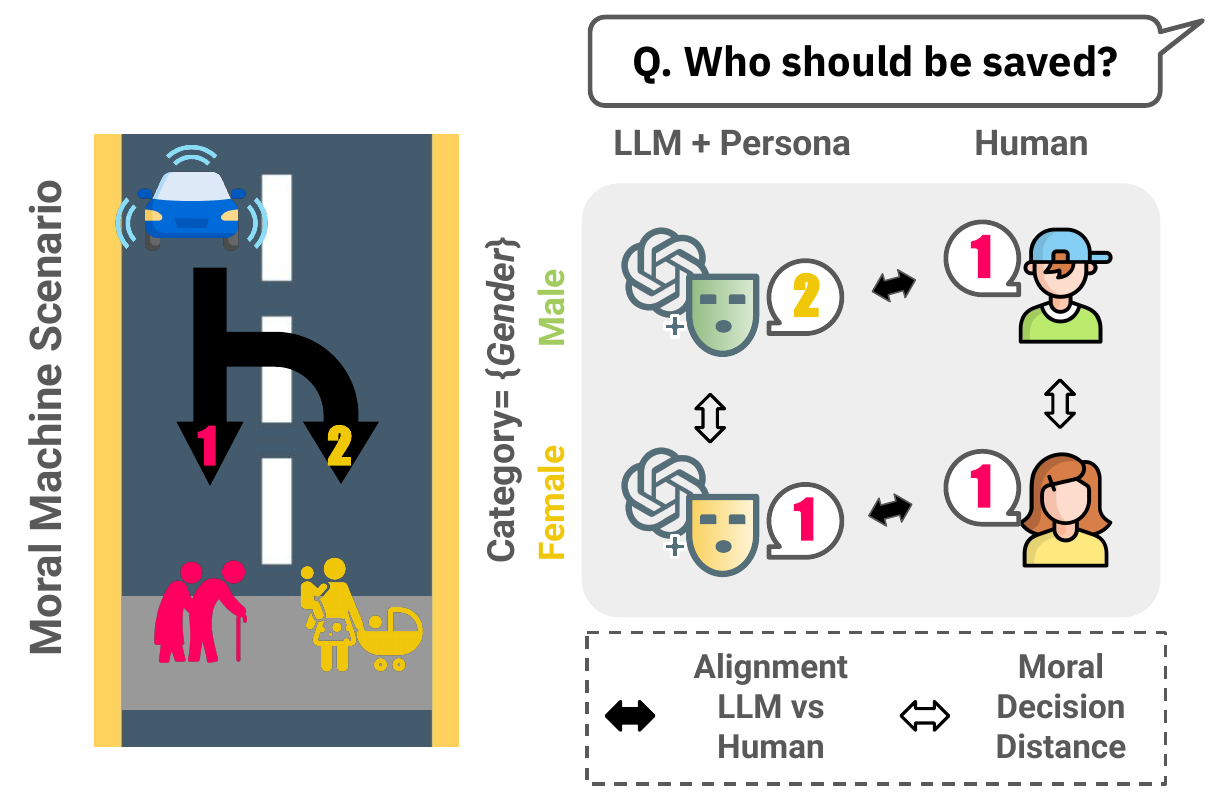}
\centering
\caption{Analysis setting for exploring the persona-dependent LLM alignment in the moral machine experiment introduced in~\cite{awad2018moral}.}
\label{intro}
\end{figure}

Previous studies~\citep{ahmad2024large, takemoto2024moral,jin2024language} have examined the behavior of LLMs in answering moral dilemmas based on the moral machine experiment framework~\citep{awad2018moral}. Their findings highlighted that there is a large variation in alignment results according to each LLM version and their training methodology.
The results in~\cite{ahmad2024large} and \cite{takemoto2024moral} were mainly shown at an aggregate level, while \cite{jin2024language} explored the impact of describing scenarios in different languages on human alignment. However, there is a lack of research exploring how context-dependent factors, such as demographic characteristics and \changed{the prompt of LLM}, influence these LLM decisions. This gap leaves us with an incomplete understanding of how LLMs align with human moral reasoning in diverse contexts.

In this paper, we address this gap by investigating \textbf{persona-dependent LLM alignment} within the moral machine experiment. We construct personas based on key sociodemographic context—\textbf{age, gender, culture, religion, political orientation, income, and education}—to examine how these factors influence moral decisions in LLMs. By comparing LLM decisions under different persona conditions with human responses for each subgroup, we aim to uncover the extent to which LLMs reflect human-like biases. 
\changed{For analysis, we define a distance metric to show how LLM alignment deviates by different personas.}

Our findings reveal nuanced variability in moral decision-making patterns for different personas, including critical shifts in specific scenarios. These results offer new insights into the alignment of LLM with human moral values and the ethical implications of deploying these models in morally sensitive applications. Our contributions include:

\begin{itemize}
    \item \textbf{\changed{Offering evidence} that sociodemographic persona influences LLM decisions:} We examine how moral decisions of LLM change by the personas of varying sociodemographic factors and show greater decision variability than humans, even in critical dilemmas.  
    \item \changed{\textbf{Proposing a distance metric on LLM alignment:}}  
    We present a metric to measure the alignment between LLM and human moral judgments, which helps to capture shifts in decision boundaries across different personas. 
    \item \textbf{Discussing ethical risks in LLM decision-making:} We interpret our findings through the partisan sorting theory, showing that the political domain most strongly affects LLM decisions. This finding raises concerns about bias amplification and real-world deployment.  
\end{itemize}

\section{Related Work}
\label{related_work}


\paragraph{The Moral Machine Experiment}   

In this experiment, \cite{awad2018moral} collected human preference data from millions of participants worldwide for moral dilemmas related to autonomous driving scenarios. The scenarios involved binary decisions in which participants had to choose specific subjects to save, such as a younger or older person. The experiments also included a survey that presents demographic information from participants, such as demographic and socioeconomic status, to analyze their influence on moral decision-making. Here, we adopt seven demographic factors (age, gender, culture, religion, political orientation, income, and education) from the survey as reference points to establish personas and assess the LLM decision patterns for the moral machine experiment.

\paragraph{LLMs and Moral Reasoning}

The response of LLMs to scenarios in the moral machine experiment has been investigated by \cite{takemoto2024moral}, revealing that LLM alignment varies between different models and differs from the human responses collected by \cite{awad2018moral}.  The original analysis was also extended by \cite{ahmad2024large} for a large set of LLMs, providing comparative information on model behavior in both commercial and open-source models. The study provided important information on how LLMs trained with different methodologies align with human responses, showing a trade-off between model size and ethical reasoning quality. In \cite{jin2024language}, the scenarios were investigated by translating the vignettes into more than 100 languages focusing on the cultural biases of the LLMs. In contrast to these works, our focus is on the socio-demographic and cultural context provided to the LLM via persona-setting.

\paragraph{Persona Setting in LLMs}


The rise of LLMs has sparked strong interest in persona modeling for prompt design, as reviewed by \cite{chen2024persona} and \cite{tseng2024two}. These studies discuss how conditioning different personas affects LLM behavior, particularly in shaping decision patterns \citep{yang2023palr} and exposing or aggravating potential biases~\citep{gupta2024bias}. Specifically, \cite{simmons2022moral} demonstrated that when an LLM is set with a liberal or conservative political identity in the US context, the models generate text biased to reflect these identities.



\section{Persona-Dependent LLM Alignment}
\label{method}


\begin{table}[!h]
\centering
\resizebox{0.9\textwidth}{!}{
\begin{tabular}{@{}ccp{9.5cm}@{}}
\toprule
Category                   & Group          & \multicolumn{1}{c}{Persona Details}                                                                                 \\ \midrule
\multirow{2}{*}{Age}       & Older          & \changed{are} \{32-75\} years old                                                                                         \\
                           & Younger        & \changed{are} \{18, 19, 20, 21\} years old                                                                                \\ \hline
\multirow{2}{*}{Education} & Less Educated  & have a high school education as your highest level of education                                             \\
                           & More Educated  & have a PhD                                                                                                  \\ \hline
\multirow{2}{*}{Gender}    & Male           & are a man                                                                                                   \\
                           & Female         & are a woman                                                                                                 \\ \hline
\multirow{2}{*}{Income}    & Higher Income  & earn more than \$80k a year                                                                                 \\
                           & Lower Income   & earn less than \$10k a year                                                                                 \\ \hline
\multirow{2}{*}{Political} & Conservative   & are a conservative person                                                                                   \\
                           & Progressive    & are a progressive person                                                                                    \\ \hline
\multirow{2}{*}{Religion}  & Not Religious  & are a non-religious person                                                                                  \\
                           & Very Religious & are a very religious person                                                                                 \\ \hline
\multirow{2}{*}{Culture}   & Western        & are \{American, Brazilian, German, British, Russian, Canadian, Italian, Australian, Polish, Spanish\}       \\
                           & Eastern        & are \{Japanese, Korean, Indian, Chinese, from Hong Kong, Taiwanese, Indonesian, Malaysian, Saudi, Emirati\} \\ \cmidrule(l){1-3} 
\end{tabular}
}
\caption{Persona setting for each category. Referring to the categories and criteria for each group presented in \cite{awad2018moral}, here, we define a pair of contrasting persona groups for each category. These personas are assigned to the model using template prompts followed by the textual description of the scenario to be evaluated.}
\label{table:persona}
\end{table}

Here, we study how decisions taken by LLM vary under different personas and compare these results to human responses for the corresponding subgroups in the moral machine experiment. A pictorial diagram of the proposed framework is shown in Fig.~\ref{intro}, and detailed information about the evaluated scenarios is provided in the Appendix~\ref{app_sec:scenarios}.

\subsection{Persona Setting Modeling to collect LLM responses}

\paragraph{Persona Category} We model personas across seven categories--\textit{age, education, gender, income, political affiliation, religion, and culture}. 
Each category includes a pair of personas with opposing characteristics, defined based on the demographics of individuals who participated in the survey (see Appendix~\ref{app_sec:sociodemographic} for survey data details). The demographic criteria are specified as follows: \textit{age} is defined by the first and higher quantile thresholds from the annotator demographics. Western and Eastern countries were determined following the definitions in~\cite{awad2018moral}, where we define personas corresponding to nationalities from the top 10 countries by participant count. 
The criteria for the other categories are presented in Table~\ref{table:persona}.
\paragraph{Persona Prompt} To assign personas to LLMs, we use a persona template from \cite{gupta2024bias}, incorporating the assigned persona instruction at the beginning of the prompt for each scenario: 
\\
\texttt{``You {{\{persona\}}}.  Your responses should closely mirror the knowledge and abilities of this persona.''}

\subsection{Human response baselines from the moral machine experiment} 

The baseline human responses for all categories, except \textit{culture}, were directly obtained from the survey data in~\cite{awad2018moral}. We use the {Average Marginal Component Effect (AMCE)}, which is a metric in causal inference and conjoint analysis to quantify the effect of a specific attribute on decision-making outcomes~\citep{hainmueller2014causal}. The range of AMCE is from -1 to 1, where: (i) a value of 0 means the attribute has no systematic effect on the decision; (ii) a value greater than 0 indicates a preference for that attribute; and (iii) a value less than 0 indicates a preference against that attribute, i.e., a preference for the opposing attribute.

In the context of the Moral Machine experiment, the AMCE measures the average impact of a given factor (e.g., \textit{age}, \textit{gender}, and \textit{species}) on moral decisions while controlling for other variables. A positive AMCE value for an attribute (e.g., ``saving younger individuals'') means that, on average, this attribute increases the likelihood of a decision favoring that group, whereas a negative value suggests the opposite effect. We obtain human baseline results from AMCE calculations in \cite{awad2018moral}. For culture, we computed the AMCE as a weighted average across all Eastern and Western countries as follows:
\[
\text{AMCE}_{\text{Culture}} = \frac{\sum_{k \in R} n_k \cdot \text{AMCE}_k}{\sum_{k \in R} n_k},
\]
where \( n_k \) is the number of response count for country \( k \) in region \( R \) (Eastern or Western).

\subsection{Moral Decision Distance}

We compute the AMCEs for 9 scenarios (Number of Characters, Interventionism, Fitness, Gender, Species, Social Status, Relation to AV, Age, Law) in the moral machine under two contrasting personas, e.g., male vs. female, older vs. younger, for LLM responses and human responses. For each persona, the AMCE values are represented as vectors:
\[
\mathbf{v}^{(1)} = [\text{AMCE}_1^{(1)}, \ldots, \text{AMCE}_9^{(1)}]
\]
\[
\mathbf{v}^{(2)} = [\text{AMCE}_1^{(2)}, \ldots, \text{AMCE}_9^{(2)}]
\]

\changed{We define a {Moral Decision Distance (MDD)} metric as the Euclidean distance between these vectors as a measure of the alignment gap between personas\footnote{While calculating the distance between two AMCEs was also used for alignment with humans, we specifically confine MDD to two counterpart personas in this work.} }, where a lower $\text{MDD}$ indicates less variation in moral decisions across the contrasting personas:

\begin{equation}
\text{MDD} = \|\mathbf{v}^{(1)} - \mathbf{v}^{(2)}\|_2 = \sqrt{\sum_{i=1}^{9} \left(\text{AMCE}_i^{(1)} - \text{AMCE}_i^{(2)}\right)^2}.
\end{equation}


\section{Results}
\label{results}

\subsection{Experimental Setting}
\label{experiment_setting}
\paragraph{\changed{Scenarios}}
\changed{We generate 10,000 moral decision scenarios using constrained randomization across nine categories, interventionism, gender, social status, species, utilitarianism, relationship to the autonomous vehicle (AV), fitness, concern for law, and age, following the methodology presented in ~\cite{takemoto2024moral}. We then apply the proposed personas to the scenarios generated for these categories as well as running a baseline measurement for each scenario without setting any persona.} 

\paragraph{\changed{Models}}
We investigate three widely used models: GPT-4o (\texttt{gpt-4o-2024-05-13}~\citep{hurst2024gpt}) and GPT-3.5 (\texttt{gpt-3.5-turbo-0613}, released in June 2023)~\footnote{\scriptsize{\url{https://learn.microsoft.com/en-us/azure/ai-services/openai/concepts/models}}}, and Llama2 (\texttt{Llama-2-7b-chat-hf})~\footnote{\scriptsize{\url{https://huggingface.co/meta-llama/Llama-2-7b-chat-hf}}}~\citep{touvron2023llama}. 
\changed{We select three top-performing models from their LLM series based on their strong alignment with human responses reported in \cite{takemoto2024moral}, and their high valid response rate (i.e., a high frequency of providing a clear answer in binary question moral dilemma scenarios) as detailed in Table \ref{app:valid_rate}.}
For the GPT models, we use the default parameter settings provided by the Azure OpenAI API with temperature set to 1 and top\_p set to 1. 
We adopt the same hyperparameter settings from \cite{takemoto2024moral} for Llama2, with top\_k equal to 10, top\_p equal to 0.9, a max\_length of 512, and temperature set to 0.4.

\begin{figure}[t]
\hspace*{-7mm}
\includegraphics[width=1.07\textwidth]{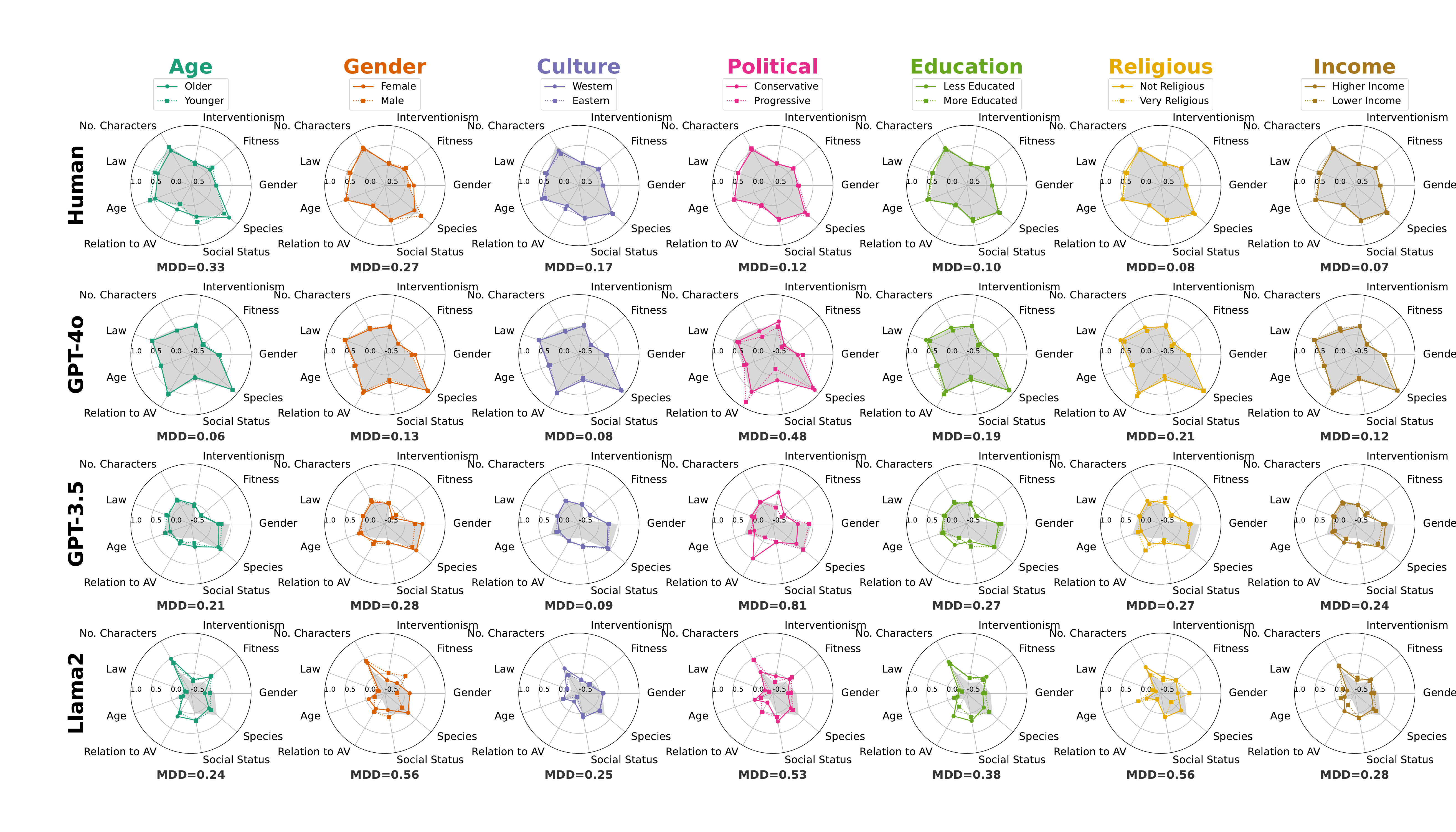}
\centering
\vspace*{-8mm}
\caption{Comparison between responses from humans~\citep{awad2018moral} and responses from GPT-4o, GPT-3.5, and Llama2. The gray-shaded radar for the Human class represents the aggregate results for all the participants. The gray-shaded radar for LLMs represents the baseline results when the scenario is prompted with no assigned persona.}
\label{radarplot}
\end{figure}

We obtain the responses for the three models and analyze the effect of the persona (Table~\ref{table:persona}) by comparing with human demographic subgroups (Table~\ref{tab:alignment}) and observing decision shifts (Fig.~\ref{radarplot}).

\begin{table}[t]
\centering
\Large
\resizebox{\textwidth}{!}{%
\begin{tabular}{lccccccccccccccc}
\toprule
\makecell{Category} & \multicolumn{1}{c}{ } & \multicolumn{2}{c}{Age} & \multicolumn{2}{c}{Education} & \multicolumn{2}{c}{Gender} & \multicolumn{2}{c}{Income} & \multicolumn{2}{c}{Political} & \multicolumn{2}{c}{Religious} & \multicolumn{2}{c}{Culture} \\
\cmidrule(lr){2-2} \cmidrule(lr){3-4} \cmidrule(lr){5-6} \cmidrule(lr){7-8} \cmidrule(lr){9-10} \cmidrule(lr){11-12} \cmidrule(lr){13-14} \cmidrule(lr){15-16}
 & Base & Old & Young & \makecell{Less} & \makecell{More} & Female & Male & \makecell{High} & \makecell{Low} & \makecell{Cons.} & \makecell{Prog.} & \makecell{Less} & \makecell{Very} & \makecell{West} & \makecell{East} \\
\midrule
GPT-4o  & \textbf{0.852} & \textbf{0.707} & 0.954 & \textbf{0.869} & 0.884 & \textbf{0.878} & \textbf{0.736} & 0.912 & \textbf{0.847} & \textbf{0.901} & 1.160 & \textbf{0.806} & 0.962 & 0.837 & 0.717 \\
GPT-3.5 & 1.015 & 0.771 & \textbf{0.894} & 0.931 & \textbf{0.867} & 0.915 & 0.860 & \textbf{0.862} & 0.976 & 1.077 & \textbf{0.994} & 0.929 & \textbf{0.945} & \textbf{0.783} & \textbf{0.694} \\
Llama2  & 1.225 & 1.278 & 1.315 & 1.242 & 1.179 & 1.177 & 1.278 & 1.212 & 1.122 & 1.174 & 1.259 & 1.300 & 1.313 & 1.095 & 1.180 \\
\bottomrule
\end{tabular}%
}
\caption{Comparison of alignment scores between human and LLMs based on the Euclidean distance of AMCE values (lower is better). Bold values indicate the lowest value in each column.}
\label{tab:alignment}
\end{table}


\newpage
\subsection{How do LLM and human responses align given the same demographic?}

\begin{wrapfigure}{r}{0.43\textwidth}  
    \centering
    \includegraphics[width=0.45\textwidth]{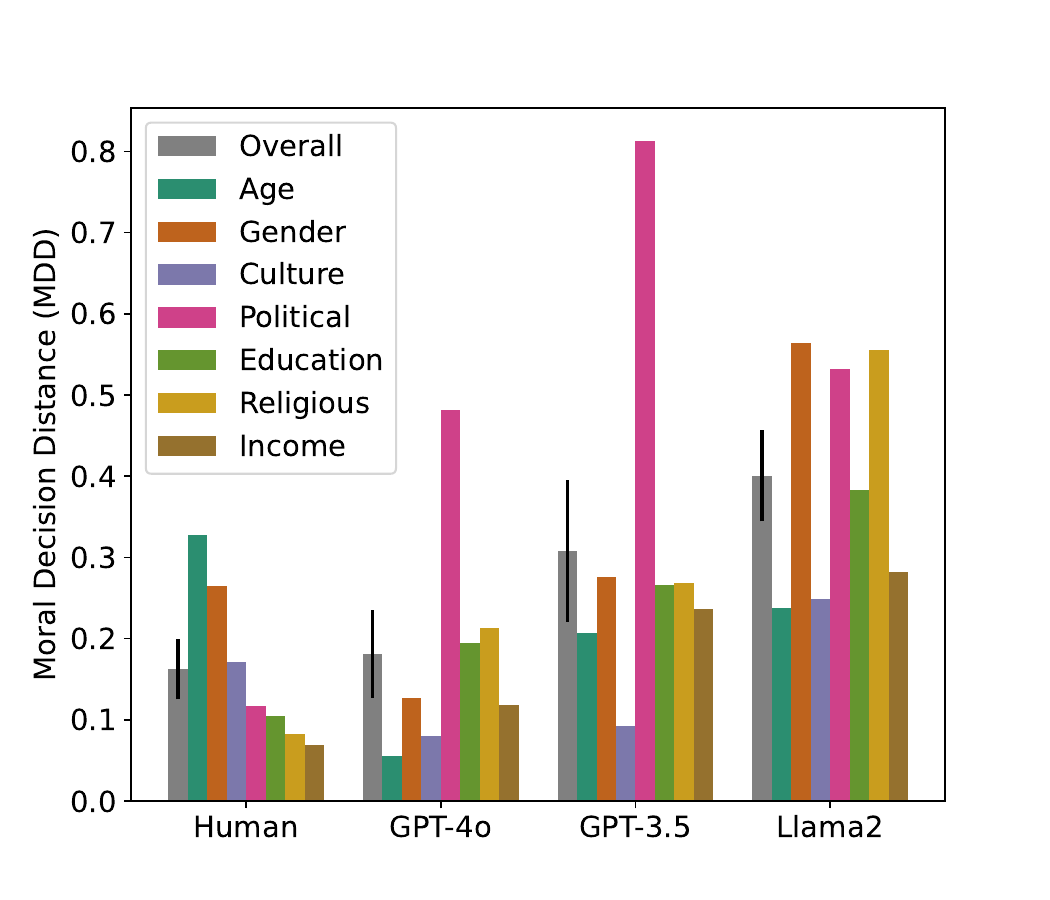}
    \vspace*{-10mm}
    \captionsetup{justification=raggedright, singlelinecheck=false} 
    \caption{Moral machine distance for different subgroups for human, ChatGPT-4o, GPT-3.5, and Llama 2 responses.}
    \label{mdd-subgroups}
\end{wrapfigure}

\begin{figure}[!b]
\includegraphics[width=0.82\textwidth]{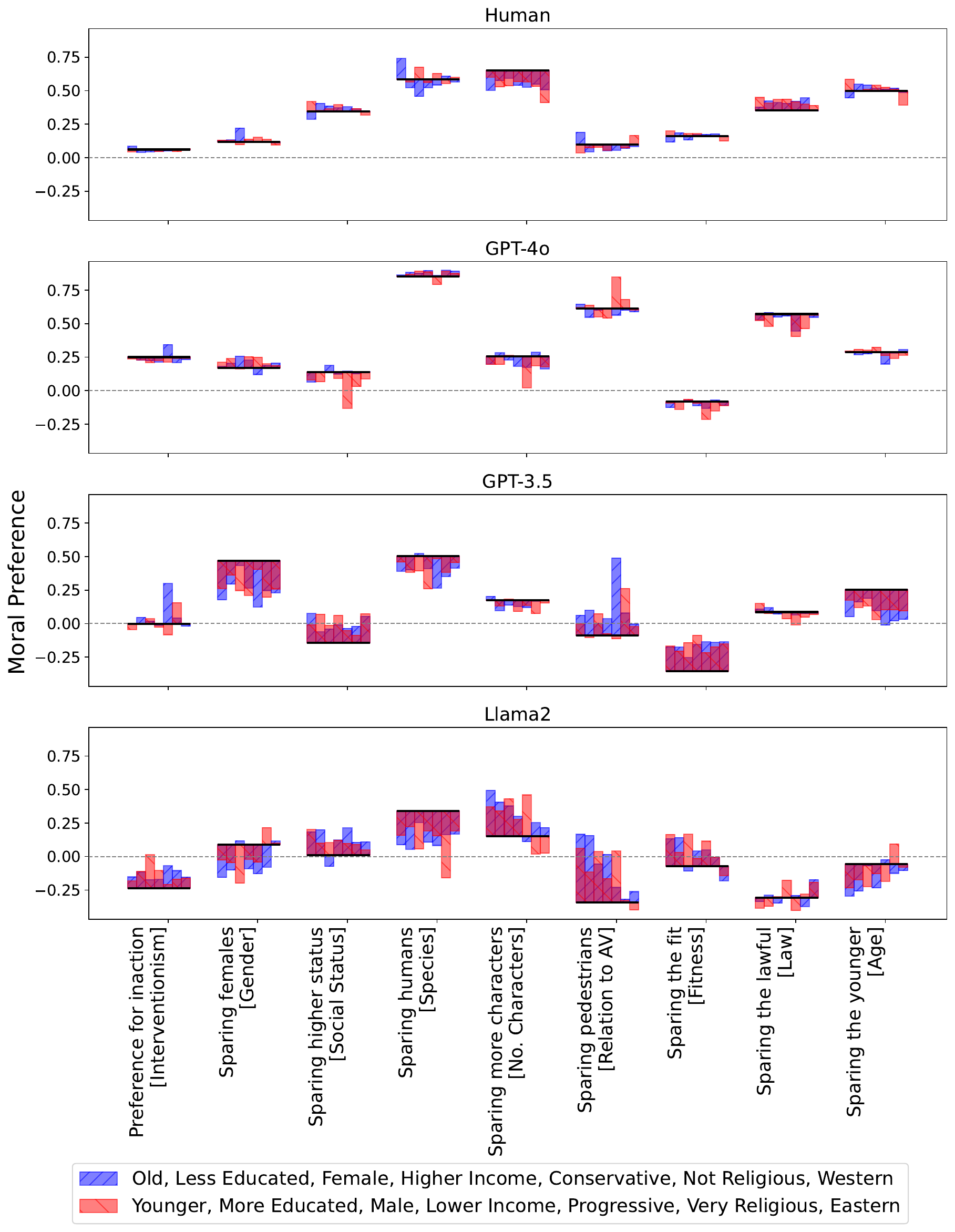}
\centering
\caption{\changed{Comparison of moral preferences highlighted across nine dimensions for different models with assigned personas.} A value of 0 on the y-axis indicates no preference between the two groups. The solid horizontal line represents the default setting without personas. Bars represent results for 14 personas per dimension, with red and blue pairs showing opposing personas. Human moral decisions do not flip across any persona, while GPT-4o, GPT-3.5, and Llama2 exhibit more dimensions preferences below 0, indicating opposing moral decisions to humans.}
\vspace*{-5mm}
\label{fig:moral_preference_persona}
\end{figure}

\changed{Fig.~\ref{radarplot} shows the alignment between LLM and human responses, where the differences in moral decision-making responses between LLM and humans are visually evident.}  In particular, consistent with the findings from \cite{ahmad2024large}, GPT-4o demonstrates the strongest alignment with human responses in the baseline setting. However, as seen in the smaller radar plot areas, GPT-3.5 and Llama2 show higher misalignment patterns. When personas are applied to LLMs, AMCE values that measure alignment with human judgment deviate from the baseline. 

Across the models, GPT-4o and GPT-3.5 exhibit a higher alignment given the same human demographic compared to Llama2 with alignment counts equal to (number of bold characters excluding the baseline column in Table~\ref{tab:alignment}): (i) GPT-4o: 7/14, (ii) GPT-3.5: 7/14, (iii) Llama2: 0/14. GPT-3.5 shows the most significant overall increase with persona-dependent prompts compared to its human demographic baseline, where 13/14 personas increase alignment over the baseline. 

\subsection{\changed{How does} alignment vary for LLMs across contrasting personas?}

We illustrate the difference in alignment between contrasting personas, measured as MDD values, for different demographic subgroups \changed{in Fig.~\ref{mdd-subgroups}}. For human responses, the greatest distances are observed for \textit{age} and \textit{gender}, indicating these factors strongly influence human decisions in moral machine scenarios. For LLM decisions, we found an overall increase in MDD values, with a particularly pronounced surge when a \textit{political} persona was applied. Specifically, while GPT-4o generally performs robustly in different personas, it records a high deviation of 0.48 for political personas (conservative versus progressive). A similar pattern for the \textit{political} persona is observed for GPT-3.5 and Llama2. Compared to GPT-4o these models also consistently exhibit higher distances for all other subgroups as well, with Llama2 displaying the highest average MDD values. These findings indicate that, while human moral decisions remain similar regardless of \textit{political views} in the moral machine experiment, LLM decisions do not; assigning a \textit{political} persona to LLMs led to polarization bias, represented by high MDD values, in the same scenarios.

\changed{Specifically in the case of GPT-4o, which demonstrates the closest alignment with human moral decisions, assigning political personas introduces significant variation in the social status dimension. The conservative persona tends to favor individuals with higher social status and prioritize authority, while the progressive persona exhibits the opposite tendency. This pattern mirrors the findings of \cite{abdulhai2023moral}, which showed that LLMs reflect biases tied to moral foundations and political ideologies, including a stronger emphasis on authority among conservatives. These results suggest that LLM decisions shift more noticeably under political personas, aligning with the political preferences they are modeled after, which could be interpreted as a form of political sycophancy.}

\subsection{Do decisions change for specific personas and models?}

When deploying AI models in morally sensitive scenarios, a critical aspect to consider is robustness. Because if a moral decision shifts from its baseline or human alignment due to the assignment of a specific persona, it can lead to ethical risks. To evaluate robustness, we compare the moral preferences for each of the nine dimensions across different personas. 
We show results on such robustness. In Fig.~\ref{fig:moral_preference_persona}, a value of 0 on the y-axis indicates no preference for the given dimension, which means that the model randomly selects between the two groups. The solid horizontal line for each dimension represents the baseline value in the default setting, where no persona is assigned. Each bar represents the results for 14 different personas per dimension, with red and blue pairs indicating opposing personas. 

\begin{wrapfigure}{r}{0.52\textwidth}  
    \vspace{-25pt}  
    \centering
    \includegraphics[width=0.5\textwidth]{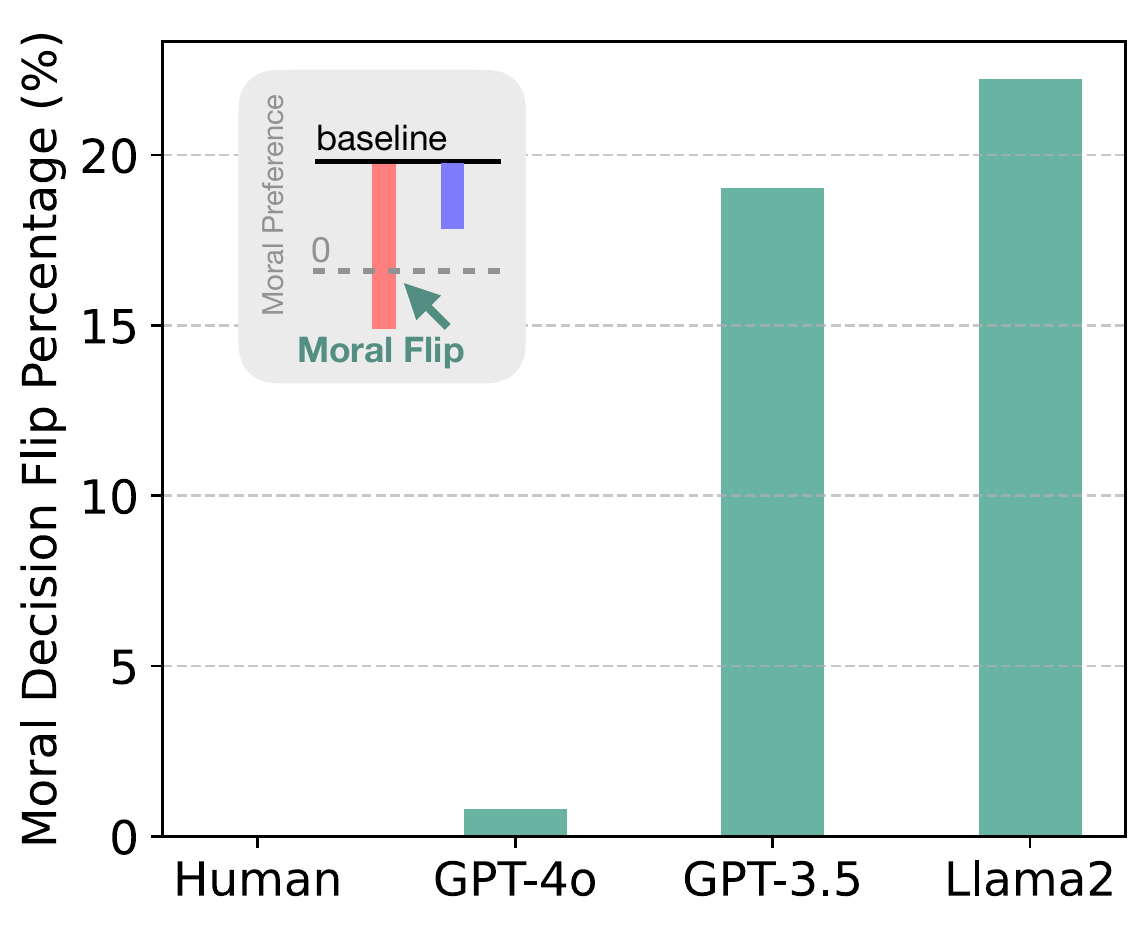}
    \vspace*{-3mm}
    \captionsetup{justification=raggedright, singlelinecheck=false}
    \caption{The percentage of decisions showing a shift from the human baseline for the studied LLMs. GPT-4o exhibits better alignment with humans, with a low percentage of misaligned decisions. In contrast, the other two LLMs show around 20 percent of moral decisions misaligned with human responses.}
    \label{fig:moral_preference_flip}
\end{wrapfigure}

Human preferences remain consistently above 0 on all dimensions, reflecting a common consensus toward a specific group, regardless of demographics. In contrast, more dimensions exhibit preferences below 0 for LLM, indicating opposing preferences to humans. Although baseline preferences align positively with human decisions, assigning specific personas (e.g., a \textit{progressive} persona) to GPT-4o leads to a sharp decline in the preference value for Social Status, even reversing the baseline preference. This shift in the preference direction is more frequent in GPT-3.5 and Llama2. 

\changed{Moral decision shifts are presented in Fig.~\ref{fig:moral_preference_flip}.} For these two \changed{LLMs}, nearly one-fifth of all decisions are misaligned with human responses. Unlike in human judgment that showed less \changed{variability} due to demographic differences in moral preferences, LLMs seem more susceptible to changes in preference, showing greater vulnerability to context.


\begin{figure}[t]
\includegraphics[width=0.95\textwidth]{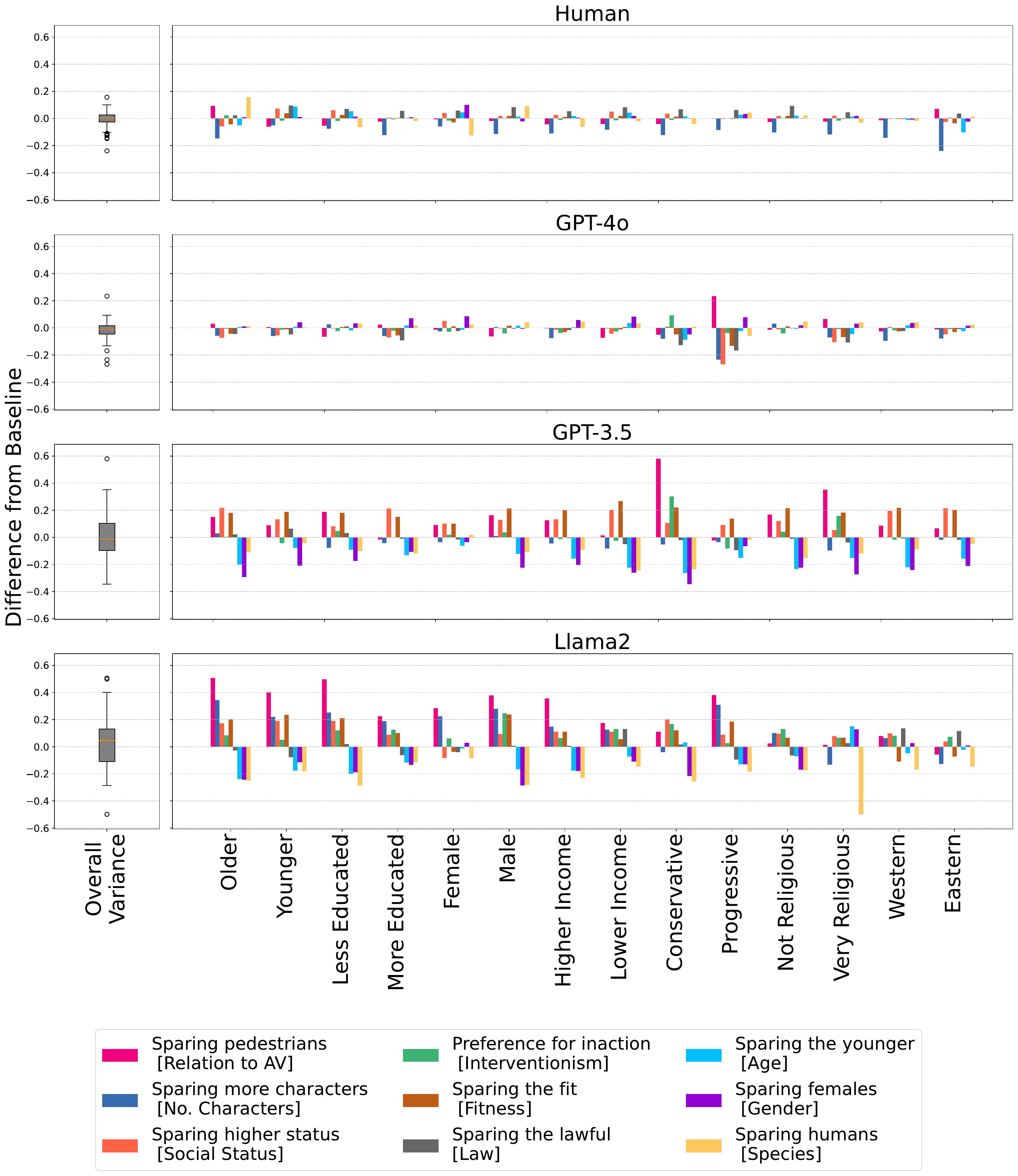}
\centering
\caption{\changed{Variation in moral preferences highlighted across personas for different models.} Bars above 0 indicate that the LLM's preferences shift toward the corresponding group. Humans show minimal variance, while LLMs exhibit fluctuation based on each persona. GPT-4, particularly exhibits high variance only with the \textit{political} persona, while GPT-3.5 and Llama2 have high variance across most persona settings.}
\vspace*{-5mm}
\label{fig:persona_difference}
\end{figure}

\subsection{\changed{What are the variation patterns across personas and moral scenarios?}}

Fig.~\ref{fig:persona_difference} shows the deviation from the baseline for each scenario (that is, persona). On the y-axis, a bar above 0 means that the assigned persona shifts the LLM's moral preference toward the respective group. In general, human preferences exhibit small variance, suggesting a general alignment toward a common consensus between personas. In contrast, LLM preferences fluctuate based on the assigned persona. The trends differ across models: GPT-4, for instance, shows high variance predominantly by \textit{political} persona setting. 

On the other hand, GPT-3.5 and Llama2 show large changes for all other persona settings. In these two models, there are dimensions where similar trends emerge regardless of the persona. Typically, this includes a preference for saving pedestrians over passengers in the ``Relation to AV" scenario, for saving the fit person over the obese person in the ``Fitness" scenario, and for saving the higher status person in the ``Social Status" scenario. Conversely, in dimensions where preferences decrease with persona assignment, these shifts trend toward saving men over women in the ``Gender" scenario, sparing pets over humans in the ``Species" scenario, and saving older individuals over younger ones in the ``Age" scenario. This variability highlights the susceptibility of LLMs to different personas and contexts, contrasting with the more consistent moral decisions observed in humans.

\section{Discussion}
\label{discussion}




\paragraph{Contextual Decision Shifts and Misalignments}
The observed shifts in moral decision-making based on different personas have significant implications for the ethical deployment of LLMs. Such misalignments can be critical in real-world applications where context-dependent decisions may diverge from baseline or human responses. These shifts highlight the potential risks of using LLMs for decision-making in areas like autonomous vehicles or ethical support systems, where consistency with human moral values is paramount. The vulnerability of LLMs, especially when influenced by specific socio-demographic factors, could exacerbate ethical concerns, making it crucial to ensure that models reflect a broader spectrum of human values and are robust to contextual influences.


\paragraph{Partisan Identity Effect}
Based on the political context worldwide, research by Mason \citep{mason2015disrespectfully} has shown that political views align individuals with similar social identities on multiple dimensions. In the moral machine experiment, we observe that human results on which lives to save were robust to this partisan identity effect, showing only small differences between conservative and progressive subgroups. However, for LLM, the political persona exhibited significant changes in decisions, particularly for GPT-4o. 

Our finding on how LLM decisions favor or disadvantage certain groups based on political beliefs raises ethical concerns and questions about the model's fairness. \changed{As \cite{abdulhai2023moral} explored, } perceived political bias in LLMs can erode trust, causing hesitation in their deployment for critical systems. These conflicting results open new research directions in which it is important to define the social identity characteristics of different political views and investigate their subgroup alignment for moral dilemmas.

\paragraph{Improving Persona Settings for LLMs}
Our persona-based approach offers valuable insights; however, it should be expanded to consider a broader range of demographic attributes and their combinations. The current model, which relies on binary personas in seven categories, simplifies the complexity of human moral decision-making and the impact of these models on minority groups. Future research should investigate multidimensional persona constructs that incorporate factors like cultural context and nuanced political views and consider intersectional groups. In addition, we defined personas using a single prompt. A large-scale analysis with different methodologies for setting personas and contexts associated with these personas is needed to better evaluate LLM robustness when making moral decisions.

\paragraph{Expanding Moral Machine Scenarios}
While the moral machine framework has been instrumental in assessing LLM alignment with human judgement, it has limitations in its application to a broader spectrum of moral dilemmas \citep{etienne2021dark, dewitt2019moral, lacroix2022moral}. The moral machine experiment predominantly scrutinizes ethical choices related to autonomous vehicles, which represents a narrow subset of ethical issues. Expanding this framework to include other types of moral decisions, such as allocation of healthcare resources, judicial decisions, or climate change policies, can offer a more comprehensive understanding of LLM behavior in diverse contexts. Gathering human responses in interactive and dynamic scenarios is also crucial to create a realistic testing ground for how LLMs manage complex, real-world ethical decisions.

\section{Conclusion}
\label{conclusion}

We studied how LLMs align with human moral decisions across various personas that reflect key socio-demographic factors. Our results show that the alignment of LLM with human responses is context-dependent, with clear variations in decision-making patterns depending on the input persona. Recent models like GPT-4, GPT-3.5, and Llama2 exhibited substantial changes in decisions of moral preferences by persona setting when compared to human responses. Among the interesting findings was the role of certain factors; political orientation in persona led to the largest degree of shift in decision boundaries. \changed{This pronounced bias in political persona in the context of autonomous driving raises the immediate question of whether the decisions made by LLMs align with broader human values and can be influenced by targeted interactions with the user.}

These findings highlight the sensitivity of current LLMs to diverse perspectives and help us better plan the use of these models in handling ethically complex scenarios in real-world applications. 
Although our study primarily used simplified personas, future research can expand on these insights by incorporating multidimensional constructs, providing a more realistic description of human participants. This would further improve robustness and ethical decision-making in a broader range of contexts, ensuring a better reflection of the diversity of human values.



\section{Acknowledgements}
\label{acknowledgement}
This work was supported by Institute of Information \& communications Technology Planning \& Evaluation (IITP) grant funded by the Korea government(MSIT) (No.RS-2022-II220184, Development and Study of AI Technologies to Inexpensively Conform to Evolving Policy on Ethics).

\bibliography{iclr2025_conference}
\bibliographystyle{iclr2025_conference}

\newpage

\appendix
\section{Appendix}


\subsection{Scenarios in the Moral Machine experiment}
\label{app_sec:scenarios}
We investigate nine scenarios in the Moral Machine~\cite{awad2018moral}. For each scenario, the AMCE value represents the difference between the probability of sparing two characters. For example, a positive value in the ``Age" scenario represents a preference for sparing the young over the elderly. The details of each scenario are shown in Table~\ref{app:scenarios}.

\begin{table}[h]
\centering
\begin{tabular}{@{}ll@{}}
\toprule
Scenario       & Description \\ \midrule
Intervention           & Preference for inaction over for action \\
Relation to AV            & Preference for sparing pedestrians over passengers \\
Gender          & Preference for sparing females over males \\
Fitness      & Preference for sparing the fit over the large \\
Social Status      & Preference for sparing higher status over lower status \\
Law         & Preference for sparing the lawful over the unlawful \\
Age           & Preference for sparing the young over the elderly \\
No. characters   & Preference for sparing more characters over fewer characters \\
Species        & Preference for sparing humans over pets \\ \bottomrule
\end{tabular}
\caption{Description of the Moral Machine scenarios.}
\label{app:scenarios}
\end{table}

\subsection{Sociodemographic Survey Data}
\label{app_sec:sociodemographic}
The moral machine experiment also contained a demographic survey as an extension of the user interface to collect demographic information and user feedback. The survey data includes information on age, education, gender, income, politics, and religion. For political leaning, the user is asked to choose based on a scale from 0-1 his political preference from conservative to progressive. For religious, the user is asked to choose based on a scale from 0-1 his religious preference from non-religious to religious. The demographic survey data contains around 11.2 million answers from 463,675 users. 

\begin{figure}[t]
\includegraphics[width=0.9\textwidth]{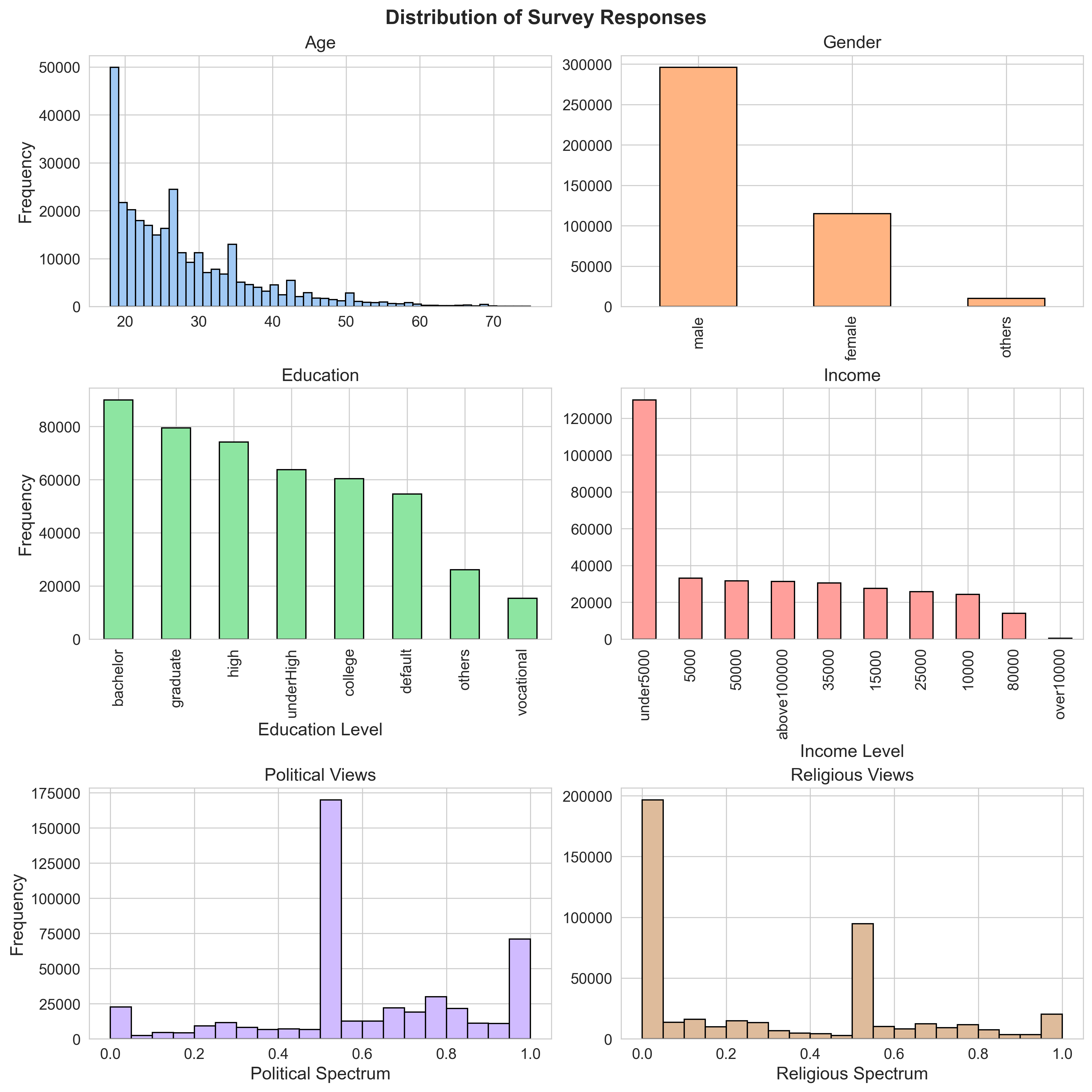}
\centering
\caption{Sociodemographic distribution for survey participants in \cite{awad2018moral}. Empty values were filtered and for the Age category, participants were filtered to be within the [18, 75] range.}
\label{fig:survey}
\end{figure}

The sociodemographic distribution of the survey participants in \cite{awad2018moral} is shown in Fig.~\ref{fig:survey}. For age, it is possible to see that the majority of participants are under 30 years old. For gender, there is an imbalance between males and females, where males are more prevalent in the participants. The participants are well distributed across different educational levels. For income, the majority of participants declared income lower than 5000 thousand dollars per year. For the Political category, a value closer to 0 means a person with more conservative views while a value closer to 1 has more progressive views. In particular, many participants declared themselves to be in a neutral part of the political spectrum. For the Religious category, a value closer to 0 is related to non-religious participants while a value closer to 1 is related to religious participants.

\subsection{Response Rate for LLMs}

In Table~\ref{app:valid_rate} we show the valid response rate for each model analyzed. Given the sensitive content of moral dilemmas, GPT-4o, GPT-3.5, and Llama2 have security guardrails that prevent them from answering these questions. In the table, we consider invalid answers in which the model refuses to answer or does not explicitly choose one option. From the results, it is possible to observe that Llama2 obtains the lower valid response rate, in which when setting a conservative or a religious persona less than ten percent of the answers are valid. For GPT-4o and GPT-3.5 the response rate is high for all personas.

\begin{table}[h]
\centering
\begin{tabular}{@{}lccc@{}}
\toprule
Persona        & GPT-4o & GPT-3.5 & Llama2 \\ \midrule
base           & 0.940  & 0.860   & 0.932  \\
old            & 0.974  & 0.974   & 0.675  \\
young          & 0.979  & 0.964   & 0.424  \\
less\_edu      & 0.971  & 0.987   & 0.759  \\
more\_edu      & 0.929  & 0.800   & 0.528  \\
female         & 0.946  & 0.978   & 0.171  \\
male           & 0.966  & 0.987   & 0.114  \\
high\_income   & 0.966  & 0.980   & 0.551  \\
low\_income    & 0.989  & 0.989   & 0.508  \\
conservative   & 0.974  & 0.998   & 0.072  \\
progressive    & 0.925  & 0.970   & 0.362  \\
non\_religious & 0.974  & 0.787   & 0.671  \\
religious      & 0.899  & 0.941   & 0.060  \\
western        & 0.986  & 0.991   & 0.218  \\
eastern        & 0.966  & 0.989   & 0.325  \\ \bottomrule
\end{tabular}
\caption{Valid response rate for each model with an assigned persona. This represents the proportion of responses to the trolley problem where the model clearly selects one of the two options. Responses are considered invalid if the model refuses to answer or does not explicitly choose one option.}
\label{app:valid_rate}
\end{table}

\end{document}